\documentclass[aps,prd,reprint,floatfix,superscriptaddress,lengthcheck,sort&compress,letterpaper,nofootinbib]{revtex4-1}

\usepackage{amssymb}
\usepackage[intlimits]{amsmath}
\usepackage{amstext} 
\usepackage{amsfonts}
\usepackage{dsfont}
\usepackage{float}
\usepackage{slashed}
\usepackage{subfigure}
\usepackage{natbib}
\usepackage{multirow}
\usepackage{verbatim}
\usepackage{nicefrac}
\usepackage{dcolumn}
\usepackage{units}
\usepackage{bm}
\usepackage{wasysym}
\usepackage{array}
\usepackage{graphicx}
\usepackage[usenames]{color}
\usepackage[colorlinks=true,linkcolor=blue,citecolor=blue,urlcolor=blue]{hyperref}
\usepackage{tabularx}
\usepackage{makecell}

\newcommand{\conjg}[1]{\ensuremath{\hspace{1pt}\overline{\hspace{-1pt}#1\hspace{-1pt}}}\hspace{1pt}}

\def\mS{\ensuremath{\mathcal{S}}}

\def\mC{\ensuremath{\mathcal{C}}}
\def\mF{\ensuremath{\mathcal{F}}}
\def\mP{\ensuremath{\mathcal{P}}}

\begin{document}

\title{Disentangling different structures in heavy-light four-quark states}

\author{Paul C. Wallbott}
\email[e-mail: ]{paul.wallbott@physik.uni-giessen.de}
\affiliation{Institut f\"ur Theoretische Physik, Justus-Liebig Universit\"at Gie{\ss}en, 35392 Gie{\ss}en, Germany}

\author{Gernot Eichmann}
\email[e-mail: ]{gernot.eichmann@tecnico.ulisboa.pt}
\affiliation{LIP Lisboa, Av.~Prof.~Gama~Pinto 2, 1649-003 Lisboa, Portugal}
\affiliation{Departamento de F\'isica, Instituto Superior T\'ecnico, 1049-001 Lisboa, Portugal}
\author{Christian S. Fischer}
\email[e-mail: ]{christian.fischer@physik.uni-giessen.de}
\affiliation{Institut f\"ur Theoretische Physik, Justus-Liebig Universit\"at Gie{\ss}en, 35392 Gie{\ss}en, Germany}
\affiliation{Helmholtz Research Academy Hesse for FAIR (HFHF), Campus Gie{\ss}en, 35392 Gie{\ss}en, Germany}

\begin{abstract}
Models proposed to explain recently discovered heavy-light four-quark states already assume certain internal
structures, i.e. the (anti)quark constituents are grouped into diquark/antidiquark clusters, heavy-meson/light-meson
clusters (hadrocharmonium) or heavy-light meson molecules. We propose and use an approach to
four-quark states based on Dyson-Schwinger and Bethe-Salpeter equations that has the potential to discriminate
between these models. We study the masses of heavy-light $cq\bar{q}\bar{c}$ and $cc\bar{q}\bar{q}$ four-quark states 
with $q=u,d,s$ and quantum numbers $I(J^{PC})=0(1^{++}),1(1^{+-}),0(0^{++})$ and $1(0^+),0(1^+),1(1^+)$. 
We identify the dominant components of the ground states with these quantum numbers and suggest candidates for corresponding
experimental states. Most notably, we find strong heavy-light meson-meson and negligible diquark-antidiquark
components in all $cq\bar{q}\bar{c}$ states, whereas for $cc\bar{q}\bar{q}$ states diquarks are present. 
A potential caveat in the $I=0$ channels is the necessary but costly inclusion of $c\bar{c}$ components 
which is relegated to future work. 
\end{abstract}

\maketitle

In the past two decades a number of highly interesting states have been identified in the charmonium and bottomonium
energy regions that cannot be accommodated for in the conventional quark model for mesons
made of a quark and an antiquark. Since the quark model is otherwise extremely successful in predicting
spectra of heavy $Q\bar{Q}$ states (with $Q=c,b$), these exceptional states are considered to be exotic hadrons.
Some of them carry electromagnetic charge and thus may be naturally explained as four-quark states $Q\bar{Q}q\bar{q}$ ($q=u,d,s$)
with a light charged quark-antiquark pair in addition to the overall neutral $Q\bar{Q}$ component.
Thus, four-quark states are considered as promising candidates to explain the properties of these exotic hadrons, see e.g.~\cite{Esposito:2016noz,Lebed:2016hpi,Chen:2016qju,Ali:2017jda,Guo:2017jvc,Olsen:2017bmm,Liu:2019zoy} for reviews.

There is, however, no agreement on the internal structure of these four-quark states.
Model approaches usually assume some kind of internal clustering from the start. One possibility, the
\textit{hadroquarkonium} picture~\cite{Voloshin:2007dx}, suggests a heavy quark and antiquark grouped together
in a tight core surrounded by the light $q\bar{q}$ pair. This is motivated by the experimental observation
of final states with a specific charmonium state and light hadrons. The
second possibility is the clustering of constituents in \textit{diquark-antidiquark} (\mbox{$dq$-$\conjg{dq}$})
components which interact via colored forces, see e.g. \cite{Esposito:2016noz} for a review. A third
possibility, especially relevant for states close to open-charm thresholds, is the
\textit{meson-molecule} picture of arrangements into pairs of $D^{(*)}\conjg{D}^{(*)}$  mesons
that interact with each other by short- and/or long-range forces \cite{Guo:2017jvc}.

It is important to note that these possibilities are not mutually exclusive: In general, every experimental
state may be a superposition of components with a different structure and the `leading' component may be different
on a case-by-case basis. It is therefore important to develop theoretical approaches to QCD that can
deal with all these possibilities. Lattice QCD is one such approach and has made
interesting progress so far, see
\cite{Prelovsek:2010kg,Prelovsek:2013cra,Abdel-Rehim:2014zwa,Lee:2014uta,Prelovsek:2014swa,
Padmanath:2015era,Francis:2016hui,Bicudo:2017szl,Francis:2018jyb,Leskovec:2019ioa} and references therein.
Nevertheless most simulations have been performed at an exploratory level using light quarks with unphysical
large masses. Functional methods, on the other hand, have
been restricted to four-quark states with equal masses \cite{Heupel:2012ua,Eichmann:2015cra} or specific
quantum numbers \cite{Wallbott:2019dng}.

In this work we present a generalization of the functional approach to four-quark states that
has the potential to systematically address and compare heavy-light states in different flavor combinations
and with different $J^{PC}$ quantum numbers. Based on a well-studied and understood truncation of the
underlying quark-gluon interaction, we work with an approximated version of the four-body Faddeev-Yakubovsky
equations that takes into account the two-body correlations that lead to the internal clustering described
above. We apply the resulting formalism to the experimentally interesting
$cq\bar{q}\bar{c}$ hidden-charm states with quantum numbers
$J^{PC}=0(1^{++})$, $1(1^{+-})$ and $0(0^{++})$, which are carried by the
$X(3872)$ \cite{Choi:2003ue,Aaij:2013zoa}, the neutral $Z(3900)$ \cite{Ablikim:2015tbp}
and (likely) the $X(3915)$ \cite{Abe:2004zs}, respectively. Furthermore, we discuss four-quark states
with open charm $cc\bar{q}\bar{q}$ in the channels $1(0^+),0(1^+)$ and $1(1^+)$.
Currently there are no experimental candidates for these states but searches are underway.
Corresponding states in the heavier bottom-quark region received a lot of attention in recent years
since they are promising candidates for deeply bound and narrow states, see e.g.
\cite{Francis:2016hui,Eichten:2017ffp, Karliner:2017qjm,Bicudo:2017szl}. It is certainly interesting
to see whether this is still the case for the experimentally more easily accessible open-charm states.

\begin{figure}[t]
  \includegraphics[width=1\columnwidth]{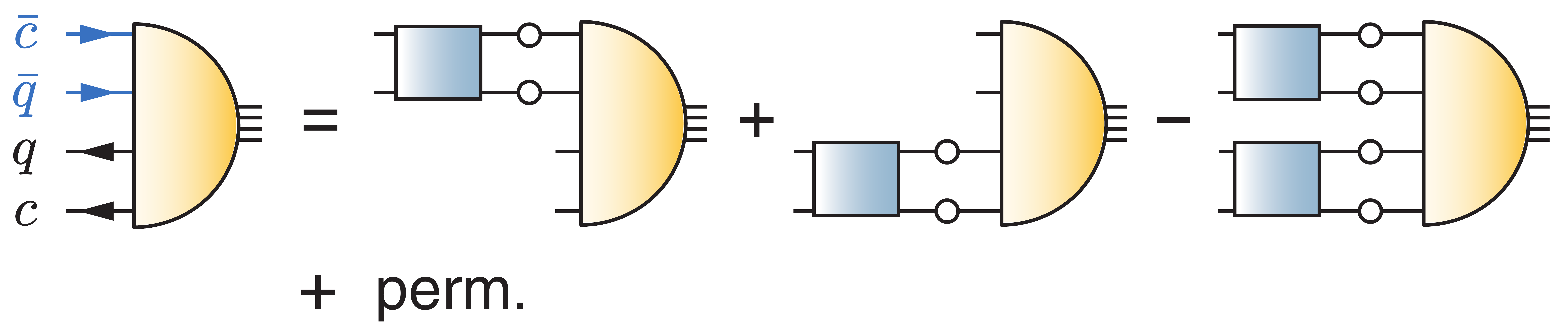}
  \caption{
  Four-quark BSE for a $cq\bar{q}\bar{c}$ system in the $(12)(34)$ configuration; the remaining $(13)(24)$ and $(14)(23)$ permutations
  are not shown. The half-circles and boxes represent the tetraquark amplitude and Bethe-Salpeter kernel, respectively.
  \label{fig_q4_bse}}
  \end{figure}

\textbf{Four-body equation.}
The homogeneous Bethe-Salpeter equation (BSE) shown in Fig.~\ref{fig_q4_bse} has the form
  \begin{equation}\label{eq_bse}
     \Gamma = K G_0 \,\Gamma \, , 
  \end{equation}
  where $\Gamma$ is the BS amplitude, $K$ is the four-quark interaction kernel that contains all possible \mbox{two-,} three- and four-body interactions,
  and $G_0$ is the product of four dressed (anti)quark propagators; see~\cite{Heupel:2012ua,Eichmann:2015cra,Wallbott:2019dng} for details.
  Each multiplication represents an integration over all loop momenta.
  Eq.~\eqref{eq_bse} holds at a given pole position of the offshell $qq\bar{q}\bar{q}$ scattering matrix $T$, which satisfies the scattering equation $T = K + K G_0\,T$.
  Poles on the real axis of the total squared momentum $P^2$ correspond to bound states,
  whereas resonances appear as poles in the complex plane on higher Riemann sheets.

\begin{table}[t]\renewcommand{\arraystretch}{1.1}
	\begin{tabular}{ c @{\;\;} | @{\;\;} c @{\;\;} | @{\;\;} c @{\;\;} c @{\;\;} | @{\;\;}c @{\;\;} c }
		& $m_{\bar{q}}$        & $m_{PS}$ & $m_V$    & $m_S$    & $m_A$    \\ \hline\hline \rule{-0.0mm}{0.35cm}
		$n\bar{n}$ &     3.7              &  138   &  732(1)  &  802(77) &  999(60) \\
		$c\bar{n}$ &     3.7              & 1802(2)  & 2068(16) & 2532(90) & 2572(8)  \\
		$c\bar{s}$ &      91              & 1911(3)  & 2169(14) & 2627(82) & 2666(7)  \\
		$c\bar{c}$ &     795              & 2792(6)  & 2980(6)  & 3382(15) & 3423(8)
	\end{tabular}
	\caption{Rainbow-ladder results for $n\bar{n}$, $c\bar{n}$, $c\bar{s}$ and $c\bar{c}$ meson and diquark masses (in MeV; $n=u,d$).
		$m_{\bar{q}}$ is the input current-quark mass at a renormalization point $\mu=19$ GeV in a momentum-subtraction scheme.
		The column $m_{PS}$ contains  the masses of $\pi$, $D$, $D_s$ and $\eta_c$,
		the column $m_V$ those of $\rho/\omega$, $D^*$, $D_s^*$ and $J/\psi$,
		and the columns $m_S$ and $m_A$ list the corresponding diquark masses. The quoted errors
		are obtained by varying the parameter $\eta=1.8 \pm 0.2$. 
		\label{tab:rl}}
\end{table}

In this work we focus entirely on the two-body correlations in $K$ since these generate the internal two-body clusters discussed above.
This leads to
\begin{equation}
KG_0 = \sum_{aa'} K_{aa'}\,, \quad K_{aa'} = K_a + K_{a'} - K_a\,K_{a'}
\label{eq_bse_kernel}
\end{equation}
where $a$, $a'$ stand for $qq$, $\bar{q}\bar{q}$ or $q\bar{q}$ pairs and $aa'$ is either $(12)(34)$, $(13)(24)$ or $(14)(23)$.
The subtraction is necessary to avoid overcounting~\cite{Huang:1974cd,Khvedelidze:1991qb,Heupel:2012ua}.
Irreducible three- and four-body interactions are not (yet) taken into account for three reasons: first, this would complicate an already
tremendous numerical task further beyond the resources currently available to us; second, a similar strategy has been employed with
great success in the baryon sector, where strong two-body correlations naturally lead to a diquark-quark picture, which in turn leads
to a spectrum in one-to-one agreement with experiment \cite{Eichmann:2016hgl,Eichmann:2016yit}; third and most important, the pictures
of internal structures that we like to discriminate (i.e. diquark/antidiquark vs. hadro-charmonium vs. meson molecule) all rely on
strong two-body clustering. Thus for the purpose of this work it is indeed sufficient to focus on two-body interactions. Nevertheless,
of course, the effects of irreducible three- and four-body forces have to be explored in future work.

For the two-body kernels we employ the same rainbow-ladder interaction that is used in the Dyson-Schwinger equation (DSE)
for the quark propagator. This truncation has recently been reviewed in~\cite{Eichmann:2016yit}, where the DSE for the quark
propagator is discussed around Eq.~(3.18) and the effective interaction
in Eqs.~(3.95--3.96). We use $\Lambda=0.72$ GeV for the scale parameter, adjusted to reproduce the pion decay
constant $f_\pi$, and $\eta=1.8 \pm 0.2$. Together with the current-quark masses, these are the only input parameters in all equations.
The construction satisfies chiral constraints such as the Gell-Mann-Oakes-Renner relation,
ensures the (pseudo-)\,Goldstone-boson nature of the pion and has been extensively applied to meson and baryon phenomenology.
As discussed in~\cite{Eichmann:2016yit}, the truncation is well-known
to reliably reproduce  many properties of pseudoscalar and vector mesons (and, correspondingly, scalar 
and axialvector diquarks). Since we focus on two-body clusters inside tetraquarks in these channels only,
we may expect qualitatively reasonable results.

The quantitative reliability of the approximation of the two-body kernel may be judged from the results for meson masses in Table~\ref{tab:rl}.
We work in the isospin symmetric limit where $m_{D^+} = m_{D^-} = m_{D^0}$. The $u/d$ current-quark mass is fixed by $m_\pi$, the strange quark
mass is chosen such that the sum $m_{D_s}+m_{D_s^*}$ equals the sum of the experimental values~\cite{Tanabashi:2018oca} and analogously
for the charm quark mass in $m_D+m_{D^*}$. The deviations between the theoretical and experimental meson masses are then
below $7\%$ in all cases.

\smallskip
\textbf{Four-quark amplitude.}
  The main challenge in solving Eq.~\eqref{eq_bse} for given $J^{PC}$ is the structure of the BS amplitude.
  Its general decomposition can be written as
  \begin{equation}\label{general-decomposition}
    \Gamma^{(\mu)}_{\alpha\beta\gamma\delta}(p_1 \dots p_4) = \sum_i f_i(\dots)\,\tau_i^{(\mu)}(p_1 \dots p_4)_{\alpha\beta\gamma\delta}\,,
  \end{equation}
  where the Lorentz-invariant dressing functions $f_i(\dots)$ depend on the ten Lorentz invariant momentum variables
  that can be constructed from four independent momenta. The tensors $\tau_i$ are the direct products of Dirac, color and flavor parts.
  A $J=0$ state has 256 linearly independent Dirac tensors and a $J=1$ state  768,
  which are collected in Ref.~\cite{Eichmann:2015cra} and the Appendix of~\cite{Wallbott:2019dng}.
  The color part of the amplitude consists of two independent
  color-singlet tensors and the flavor wave functions depend on the particular system, cf.~Appendix~\ref{app-construction} for details.

  To extract physical content from the BS amplitude $\Gamma^{(\mu)}(p_1 \dots p_n)$, we observe that the amplitude develops
  internal two-body clusters, which for heavy-light systems occur in the three different channels corresponding to
  hadroquarkonia, heavy-light meson-meson components and $dq$-$\bar{dq}$ clusters.
  These clusters may go on-shell provided that the sum of their masses is smaller than the mass of the
  four-body state. If this occurs in color-singlet channels, the four-quark state becomes a
  resonance in the two-body hadronic system of the corresponding clusters. But even if the masses of the
  two-body clusters are large enough such that the probed momenta only come close to the corresponding
  singularities, this will influence the four-body system. Thus the guiding idea is to represent $\Gamma^{(\mu)}(p_1 \dots p_n)$
  in terms of these two-body clusters.

  Since we are interested in specific quantum numbers with experimental candidates for four-quark
  states, we draw on existing information on the decay channels of these states and construct our
  representation along the content displayed in Table~\ref{tab_amplitude_content}. For example, in the heavy-light meson
  sector we took into account combinations of the $I(J^P)=1/2(0^-)$ multiplets $(D,D^*)$ and their anti-particles $(\bar{D},\bar{D}^*)$,
  omitting the heavy combination $D^{*} \conjg{D}^{*}$. Note that since we work in the isospin-symmetric
  limit the charged and neutral states are mass-degenerate and both taken into account.

  \begin{table}[t] \renewcommand{\arraystretch}{1.4}
 	\begin{tabular}{l @{\qquad} l @{\qquad} l @{\qquad} l @{\qquad} l}
                                & $I(J^{P(C)})$                 & Physical components           \\[1mm]  \hline \rule{-1.3mm}{0.4cm}		
        $cq\overline{qc}$       & $0(0^{++})$   & $DD$, $J/\psi\,\omega$, $SS$      \\
                                & $0(1^{++})$   & $DD^\ast$, $J/\psi \,\omega$, $SA$     \\
                                & $1(1^{+-})$   & $DD^\ast$, $J/\psi \,\pi$, $SA$ \\[1mm]  \hline \rule{-1.3mm}{0.48cm}	
        $cc\overline{qq}$       & $1(0^{+})$    & $DD$, $AA$     \\
                                & $0(1^{+})$    & $DD^\ast$, $AS$     \\
                                & $1(1^{+})$    & $DD^\ast$, $AA$     \\

 	\end{tabular}
 	\caption{Physical content of the BS amplitudes for flavor combinations $cq\overline{qc}$ and  $cc\overline{qq}$. Scalar and axialvector diquarks are denoted by $S$ and $A$, respectively.
 		\label{tab_amplitude_content}}
 \end{table}

  The construction of the $\tau_i^{(\mu)}(p_1 \dots p_4)$ for 
  the configurations $qq\overline{qq}$, $cq\overline{qc}$ and $cc\overline{qq}$ is detailed in App.~\ref{app-construction}.
  We construct the Dirac parts according to the dominant two-body clusters 
  and combine them with appropriate color and flavor wave functions such that the charge-conjugation and Pauli exchange symmetries are respected. 
  As a result, one populates a physically motivated subset of all possible basis elements. 
  Whereas in this work we restrict ourselves to the combinations displayed in Table~\ref{tab_amplitude_content},
  in principle one could construct a complete basis for the four-body amplitude with entangled Dirac, color and flavor tensors
  including all possible meson and diquark channels (i.e., also those with higher total angular momentum).

\begin{figure}[t]
	\includegraphics[width=0.9\columnwidth]{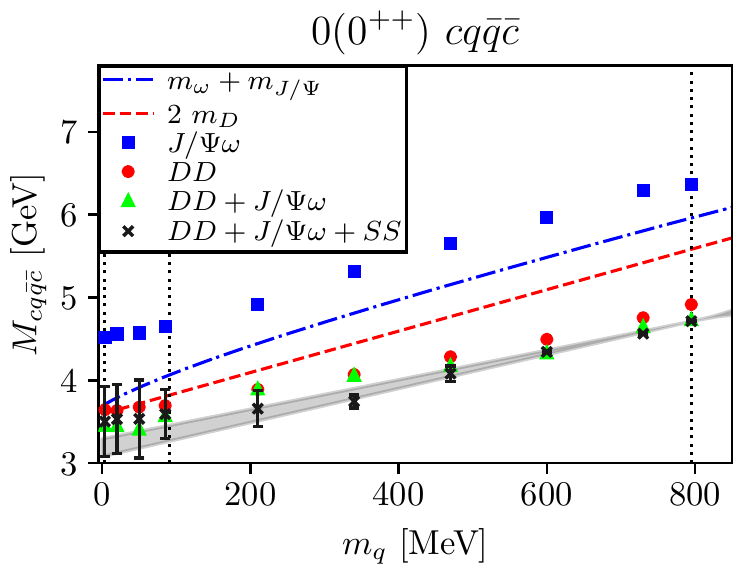} \\
    \includegraphics[width=0.9\columnwidth]{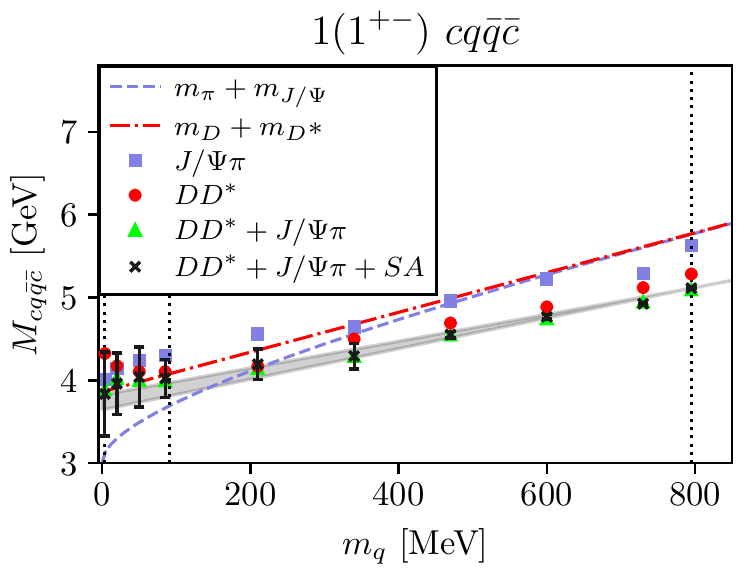}
	\caption{Quark-mass evolution of the $cq\overline{qc}$ ground states in the $0(0^{++})$ and $1(1^{+-})$ channels
             for different components of the four-body amplitude. The three vertical dashed lines mark the positions of the 
             up/down, strange and charm quark (from left to right). Masses below the respective two-meson thresholds have been 
             determined directly from the eigenvalue curve $\lambda(P^2)$ of the BSE. Results above the threshold (i.e. for small 
             quark masses) are obtained from extrapolated eigenvalue curves.
             For example, in the upper plot, the masses obtained with $J/\Psi \omega$ components only are all extrapolated, 
             whereas in the other three setups all results are read off directly, except those at the smallest quark masses 
             which lie above the lowest threshold. See main text for further details.
		      \label{fig_ZX}}
\end{figure}

\begin{figure}[t]
	\centering
	\includegraphics[width=0.9\columnwidth]{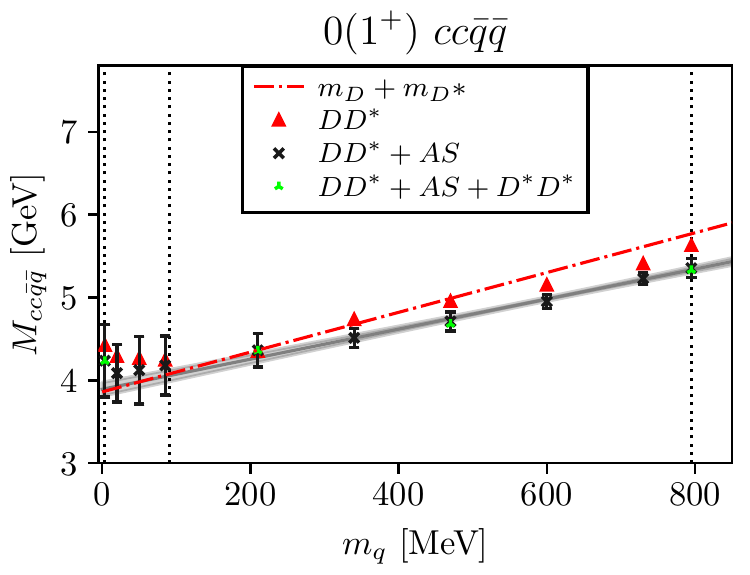} \\
	\includegraphics[width=0.9\columnwidth]{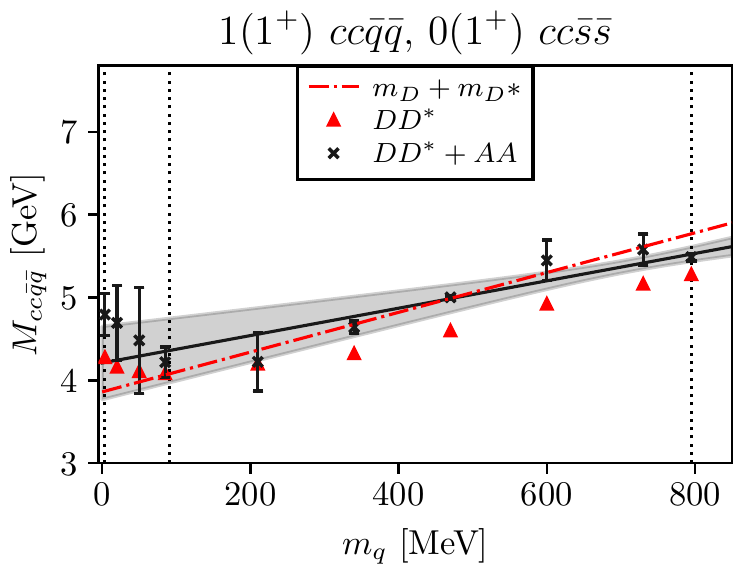}
	\caption{Quark-mass evolution of the $cc\overline{qq}$ ground states in the  $0(1^+)$ and $1(1^+)$ channels.
		The $0(1^+)$ $cc\overline{ss}$ state is read off from the curve for the $1(1^+)$ state at $m_q = m_s$. 
		The three vertical dashed lines mark the positions of the 
		up/down, strange and charm quark (from left to right). Masses below the $DD^*$ threshold have been 
		determined directly from the eigenvalue curve $\lambda(P^2)$ of the BSE. Results above the threshold (i.e. for small 
		quark masses) are obtained from extrapolated eigenvalue curves.\label{fig_ccuu}}
\end{figure}

With this setup, we are in a position to solve the BSEs as an eigenvalue problem with structure 
$\lambda \, \Gamma = K G_0 \,\Gamma$ and general eigenvalue $\lambda$. All elements of this equation depend
explicitly on the total momentum $P$. By varying $P^2$ such that $\lambda=1$ one finds the mass of the bound 
state/resonance in the four-body system via $P^2=-M^2$. 
One problem that appears in this search is the potential appearance of singularities
in the plane of the complex total momentum due to the internal meson and diquark correlations. Typically, this does not happen for large masses of the
lighter quark pair, where the resulting bound state is in general (well) below the meson-meson thresholds, cf.~Figs.~\ref{fig_ZX} and~\ref{fig_ccuu}.
However, this situation changes
for smaller masses and in particular close to the physical point of light quark masses, where the mass of the bound state/resonance comes
close to or is even larger than the meson-meson threshold. For these cases we determine the eigenvalue curve $\lambda(M^2)$
in the singularity free region and extrapolate it further into the time-like momentum domain using rational functions.
This procedure can only pick up the real part of potentially complex masses, i.e. it is not possible to extract decay widths. 
To this end one would need to use the much more involved approach described in \cite{Williams:2018adr,Eichmann:2019dts}, 
which is not (yet) at our disposal for heavy-light four-quark states. For many further details and first results in the 
light quark sector we refer the reader to Ref.~\cite{Santowsky:2020pwd}.

\smallskip
\textbf{Hidden-charm states.}
We first discuss our results for the hidden-charm $cq\overline{qc}$ four-quark states in the $I(J^{PC})=0(0^{++})$ and $1(1^{+-})$ channels.
In this context we wish to emphasise that the results in the $I=0$-channel must be seen as preliminary since the necessary but costly inclusion 
of $c\bar{c}$ components in our approach is relegated to future work (see however \cite{Santowsky:2020pwd} for 
corresponding results in the light quark sector). 

The results for the $0(1^{++})$ heavy-light state can be found in \cite{Wallbott:2019dng}, where we also described our procedure
to estimate (part of) the error of the calculation. In that case we found a dominant
heavy-light meson component, whereas the hadrocharmonium component is rather weak and the diquark component
has almost no effect at all.

\begin{table}[t] \renewcommand{\arraystretch}{1.4}
	\begin{tabular}{c @{\quad} l @{\quad} l @{\quad} | @{\quad} c @{\quad} l @{\quad} l}
		$I(J^{PC})$ & $cn\overline{nc}$     & $cs\overline{sc}$    &    $I(J^{P})$  & $cc\overline{nn}$     & $cc\overline{ss}$       \\[1mm]  \hline \rule{-1.3mm}{0.48cm}	
		$0(0^{++})$ & 3.20(11)               & 3.36(10)            &    $0(0^+)$    &  	--		             &  3.95(10)                          \\
		& 3.50(42)               & 3.59(30)            &    $1(0^+)$    & 3.80(10)		             &      --                        \\
		$0(1^{++})$ & 3.92(7) 			     & 4.07(6)             &    $0(1^+)$    & 3.90(8)                &   4.36(39)                         \\
		$1(1^{+-})$ & 3.74(9) 			     &   --                &    $1(1^+)$    & 4.22(44) 			     &    --
	\end{tabular}
	\caption{Masses of hidden-charm ($cq\overline{qc}$) and open-charm ($cc\overline{qq}$) states ($n=u,d$) in GeV.
		The combined error from fitting and varying the momentum partitioning is given in parentheses.
		For the $0(0^{++})$ state we quote both the fitted value (first line) and the direct calculation (second line).
		Tetraquarks with open or hidden strangeness are only quoted in channels where physical
		states may be present.
	}
	\label{tab_results}
\end{table}

A similar pattern arises in the scalar $0(0^{++})$ case displayed in the top panel of Fig.~\ref{fig_ZX}.
We show the mass evolution of the four-quark
state when the mass of the $c\bar{c}$ pair is fixed and the 
mass of the other $q\bar{q}$ pair is varied from the charm mass (rightmost vertical dashed line) to the strange and
light quark masses (other two vertical lines). We compare calculations with hadrocharmonium content
only ($J/\Psi \,\omega$; squares), heavy-light meson components only ($DD$; circles) and a combination of the two (triangles).
The full calculation including also the diquark/antidiquark component ($DD+J/\Psi\,\omega+SS$) is marked with crosses.
Not contained in the figure is our result for using diquark components only ($SS$). For the scalar channel and in fact for
all channels that we studied we obtain very large masses with diquarks only. These are typically
of the order of the diquark/antidiquark thresholds around 5 GeV (cf. Table \ref{tab:rl}) and therefore unphysical.
Comparing the results for $0(0^{++})$ shown in Fig.~\ref{fig_ZX} we find that for large masses, the \mbox{$dq$-$\conjg{dq}$} component
has almost no effect on the results, whereas the hadrocharmonium component gives only mild corrections
to the leading heavy-light meson components.
The diquark corrections become somewhat more prominent for small masses, however without changing the general picture.
Due to the sizeable error bar we cannot discriminate between a bound state and a resonance.
The masses quoted in Table~\ref{tab_results} are obtained from
a linear fit to the mass evolution at larger quark masses;
only for $0(0^{++})$ the direct calculation yields a mass below any threshold.

The general picture changes somewhat for the $1(1^{+-})$ state shown at the bottom of Fig.~\ref{fig_ZX}.
Once again we find negligible diquark components and a strong heavy-light meson component, but also non-negligible
contributions from the hadrocharmonium component. It is interesting to compare this behavior to the
$1^{++}$ channel discussed in \cite{Wallbott:2019dng}. From Table~\ref{tab_amplitude_content} we observe that
the only difference between the two states are the hadrocharmonium components.
The $J/\psi \,\pi$ component in the $1^{+-}$ channel is significantly lighter
than the $J/\psi \,\omega$ component in the $1^{++}$ channel, which lifts the degeneracy between the two states and
leads to a lighter mass of the $1^{+-}$ (which is, however, opposite to the current experimental situation).
We have also tested further components, which can contribute to the axialvector $1(1^{+-})$ state. We
found the $D^\ast D^\ast$ components to be negligible; however, the $\eta_c\,\rho$ component is sizeable
and enhances the mass splitting once included. Further studies in this direction are necessary.

The resulting masses are collected in Table~\ref{tab_results}. It is interesting to compare
our results with expectations from the literature. In Ref.~\cite{Cleven:2015era} heavy-quark symmetry has been
used to predict patterns for molecular states. This led to the identification of the $X(3872)$ with a molecular
state in the $1^{++}$ channel and the neutral $Z_c(3900)$ with a molecular state in the $1^{+-}$ channel. Our
results agree with this identification: for the $X(3872)$ we expect an almost pure heavy-light meson state which
is then natural to expect to sit very close to the $D\bar{D}^\ast$ threshold. For the $Z_c(3900)$, however, we find
non-negligible corrections from other components, which may shift the physical state away from the
threshold. For the scalar channel no predictions
have been made in~\cite{Cleven:2015era}, see however \cite{Gamermann:2006nm,Gamermann:2007mu,Dai:2015bcc}
for a detailed discussion. The lightest scalar molecule, if it exists, would be expected
at the $D\bar{D}$ threshold. This is indeed the case for our scalar state, which sits in the region of the
threshold of our $D$ mesons (cf. Table~\ref{tab:rl}). Thus the mass pattern emerging in Table~\ref{tab_results}
is in line with our observation of heavy-light meson dominance in all $cq\bar{q}\bar{c}$ states studied so far.

In Table~\ref{tab_results} we also list the masses of the charm-strange $cs\bar{s}\bar{c}$ states
extracted from the mass evolution.
For $I=0$ these correspond to observable states, whereas for
$I=1$ they are unphysical. 
In  the $1^{++}$ channel there is an experimental candidate, namely the $X(4140)$ with a mass only
slightly above the upper range of our error bar. Provided this identification holds, we predict a strong
heavy-light meson component of the $X(4140)$, even though it is not overly close to the $D_s\bar{D}_s^\ast$ threshold.
In addition, we find a corresponding state in the $0^{++}$ channel, although the large error bars in this
case make a prediction of its mass rather imprecise.

\smallskip
\textbf{Open-charm states.}
Our results for the open-charm states with flavour content $cc\overline{qq}$ are shown in Fig.~\ref{fig_ccuu}.
The basis construction in the open-charm case is significantly different from hidden charm
since charge-conjugation symmetry is replaced by Pauli antisymmetry.
The two heavy-light meson combinations $(c\bar{q})(c\bar{q})$ are identical,
whereas the \mbox{$dq$-$\conjg{dq}$} component $(cc)(\bar{q}\bar{q})$ with a heavy diquark and a light antidiquark
inherits the role of the hadrocharmonium component. For $I=0$,
 the light $\conjg{qq}$ antidiquark must be scalar ($S$) due to symmetry, whereas for $I=1$ it is
axialvector ($A$). The heavy $cc$ pair is always an axialvector diquark.

This change of internal
dynamics is also reflected in the results. The heavy-light meson component alone produces a state that is well
below the $DD^*$ threshold and moves only slightly above threshold for decreasing quark masses.
Whereas the $AS$ diquark contribution in the $I=0$ case is negligible compared to the
$DD^*$ contribution, the $AA$ diquark component for $I=1$ has a significant impact in pushing the mass evolution up
 above threshold.
 In the graph for $0(1^+)$ we also included the $D^\ast \bar{D}^\ast$ component explicitly, which is negligible as in all other channels.
 The resulting mass hierarchy between the isosinglet and isotriplet states
is as expected from heavy quark symmetry \cite{Eichten:2017ffp}. Our extrapolated values for the masses are in
the ballpark expected from other approaches, see e.g. \cite{Eichten:2017ffp,Luo:2017eub,Karliner:2017qjm} and Refs.
therein.

In the open-charm case, the $cc\overline{ss}$ states (with $I=0$) must be read off from the $cc\overline{qq}$ curves with $I=1$
since those have the same wave-function components cf.~Appendix~\ref{app-construction}.
As a consequence, several slots in Table~\ref{tab_results} are empty because they do not support physical states.
Moreover, there is a large gap between the light and strange state in the $0(1^+)$ channel but
it comes again with a sizeable mass uncertainty.

\smallskip
\textbf{Conclusions.}\label{sec:conclusion}
In this work we have studied and compared the masses of heavy-light four-quark states in the charm energy region.
We developed a dynamical framework that takes into account all possible combinations of internal two-body
clusters and is therefore able to decide dynamically whether pictures from effective field theory and models
(meson molecule, hadrocharmonium, diquark-antidiquark) are realized. For hidden charm, in all cases considered
we do not find a sizeable diquark-antidiquark component. Instead, the heavy-light meson component
is favoured, with channel-dependent negligible (for $0(1^{++})$) or small but significant ($0(0^{++})$ and $1(1^{+-})$)
contributions from the hadrocharmonium component. The situation for open charm is similar; the dominant
contribution is the heavy-light meson component although significant corrections from the diquark-antidiquark
component arise. Although the masses of the charm quarks are far from static, mass
patterns expected from heavy-quark symmetry are visible: In the open-charm sector we observe the expected
mass hierarchy and difference between the $I=0$ and $I=1$ axialvector channels~\cite{Eichten:2017ffp}.
The observed mass pattern in the hidden charm sector resembles the one expected of the lowest-lying multiplet of
states in the hadronic molecular approach \cite{Cleven:2015era}. In order to make further contact
with heavy-quark symmetry and lattice QCD, it would be interesting to further increase the masses of the heavy
quarks and explore the bottomonium sector. This is technically challenging and therefore left for future studies.
A potential caveat of the present formalism is that it does not take into account potentially important effects
from mixing with ordinary $c\bar{c}$ states in the $I=0$ channels \cite{Prelovsek:2013cra,Padmanath:2015era}.
Again, this is left for future work.

\smallskip\newpage
\textbf{Acknowledgments.}
We are grateful to Christoph Hanhart, Soeren Lange, Sasa Prelovsek and Marc Wagner for discussions.
This work was supported by the DFG grant FI 970/11-1 and by the FCT Investigator Grant IF/00898/2015.

\bibliographystyle{apsrev4-1}
\bibliography{X_tetraquark2}

   \newpage
   
\begin{appendix}

 \begin{figure}[t]
  \includegraphics[width=0.8\columnwidth]{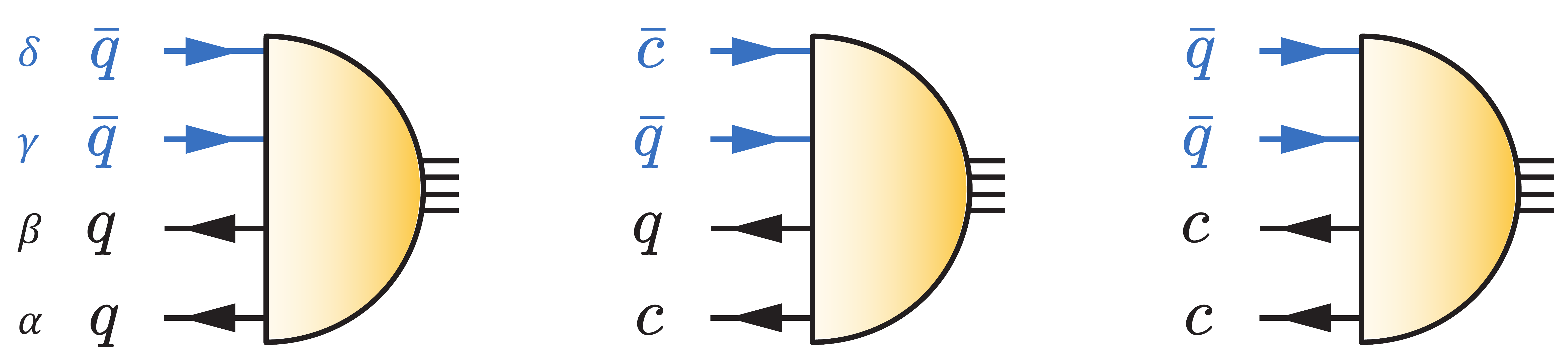}
  \caption{Examples of four-body amplitudes with different quark content: $qq\overline{qq}$ (identical quarks), $cq\overline{qc}$ (hidden charm) and $cc\overline{qq}$ (open charm).
  \label{fig_3confs}}
  \end{figure}

\section{Construction of the four-quark Bethe-Salpeter amplitude} \label{app-construction}

  In this appendix we describe the construction of the tetraquark amplitude $\Gamma^{(\mu)}_{\alpha\beta\gamma\delta}(p_1, p_2, p_3, p_4)$ in detail.
  It is the direct product of Dirac, color and flavor parts; here, $\{ \alpha, \beta, \gamma, \delta \}$
  either represent the Dirac indices or absorb all Dirac, color and flavor indices to avoid clutter.
  If we consider the color parts alone, we work with latin indices $\{A,B,C,D\}$.
  We also suppress the Lorentz index $\mu$ that distinguishes between $J=0$ and $J=1$ states.

  We consider the three different physical systems shown in Fig.~\ref{fig_3confs}:
  a state with four identical quarks $qq\overline{qq}$,
  a hidden-charm system $cq\overline{qc}$,
  and an open-charm state $cc\overline{qq}$.
  Depending on how we connect the indices, each system
  can come in a diquark-antidiquark (\mbox{$dq$-$\conjg{dq}$}) configuration (12)(34) and two meson-meson configurations (13)(24) and (14)(23).
  In terms of diquark and meson clusters,
  the hidden-charm system $cq\overline{qc}$ contains
  two heavy-light diquarks in (12)(34), whereas (13)(24) describes a `molecule'
  of two heavy-light mesons and (14)(23) a `hadrocharmonium' with a heavy and a light meson.
  For the open-charm system $cc\overline{qq}$, (12)(34) contains a heavy diquark and a light antidiquark
  and the two meson-meson configurations are both molecular.

  Depending on which system we study, the amplitude is subject to certain symmetry constraints, namely Pauli antisymmetry in (12) or (34)
  and charge-conjugation symmetry in (13)(24) or (14)(23). Pauli antisymmetry in (12) or (34) means
  \begin{align}
     \Gamma(p_2, p_1, p_3, p_4)_{\beta\alpha\gamma\delta} \stackrel{!}{=}  &- \Gamma(p_1, p_2, p_3, p_4)_{\alpha\beta\gamma\delta}\,, \label{pauli-12} \\
     \Gamma(p_1, p_2, p_4, p_3)_{\alpha\beta\delta\gamma} \stackrel{!}{=}  &- \Gamma(p_1, p_2, p_3, p_4)_{\alpha\beta\gamma\delta}\,, \label{pauli-34}
  \end{align}
  where a permutation of all (Dirac, color and flavor) indices is understood. Likewise,
  charge-conjugation symmetry in (13)(24) amounts to
  \begin{equation} \label{cc-13-24}
  \begin{split}
     &C_{\alpha\alpha'}\,C_{\beta\beta'}\,C_{\gamma\gamma'}\,C_{\delta\delta'}\,\Gamma(-p_3,-p_4,-p_1,-p_2)_{\gamma'\delta'\alpha'\beta'} \\
     & \qquad \qquad \stackrel{!}{=} \pm \Gamma(p_1, p_2, p_3, p_4)_{\alpha\beta\gamma\delta}
  \end{split}
  \end{equation}
  and in (14)(23) to
  \begin{equation} \label{cc-14-23}
  \begin{split}
     &C_{\alpha\alpha'}\,C_{\beta\beta'}\,C_{\gamma\gamma'}\,C_{\delta\delta'} \,\Gamma(-p_4,-p_3,-p_2,-p_1)_{\delta'\gamma'\beta'\alpha'} \\
     & \qquad \qquad \stackrel{!}{=} \pm \Gamma(p_1, p_2, p_3, p_4)_{\alpha\beta\gamma\delta}\,,
  \end{split}
  \end{equation}
  where $C = \gamma_4 \gamma_2$ is the charge-conjugation matrix and the signs $\pm$ determine the $C$ parity of the state.
  The amplitude for a $qq\overline{qq}$ system satisfies all four relations, whereas
  a hidden-charm amplitude $cq\overline{qc}$ is only subject to Eq.~\eqref{cc-14-23} and
  an open-charm configuration $cc\overline{qq}$ only satisfies Eqs.~(\ref{pauli-12}--\ref{pauli-34}).

  To proceed, we write the most general tensor basis decomposition of the amplitude as
  \begin{equation}\label{general-decomposition2}
    \Gamma(p_1 \dots p_4)_{\alpha\beta\gamma\delta} = \sum_i f_i(\dots)\,\tau_i(p_1 \dots p_4)_{\alpha\beta\gamma\delta}\,,
  \end{equation}
  where the Lorentz-invariant dressing functions $f_i(\dots)$ depend on the ten Lorentz invariant momentum variables
  that can be constructed from four independent momenta. The tensors $\tau_i$ are the direct products of Dirac, color and flavor parts.

  Let us first work out the color tensors.
  From $\mathbf{3}\otimes \mathbf{3} \otimes \mathbf{\bar{3}} \otimes \mathbf{\bar{3}} =
       ( \mathbf{\bar{3}} \oplus \mathbf{6}) \otimes (\mathbf{3}\oplus\mathbf{\bar{6}}) = \mathbf{1} \oplus \mathbf{1} \oplus ... $, the
  color part of the amplitude
  consists of two independent color singlet tensors, which can be taken from
  the \mbox{$dq$-$\conjg{dq}$} ($\mathbf{\bar{3}} \otimes \mathbf{3}$, $\mathbf{6}\otimes\mathbf{\bar{6}}$)
  or either of the meson-meson configurations ($\mathbf{1} \otimes \mathbf{1}$, $\mathbf{8} \otimes \mathbf{8}$),
  for example:
  \begin{equation}\label{color-11}
  \begin{split}
     (\mC_{11})_{ABCD} &=  \frac{1}{3}\,\delta_{AC}\,\delta_{BD} \\
     (\mC_{11}')_{ABCD} &=  \frac{1}{3}\,\delta_{AD}\,\delta_{BC}\,.  
  \end{split}
  \end{equation}
  The two tensors in the \mbox{$dq$-$\conjg{dq}$} decomposition are linear combinations of these,
  \begin{equation}\label{color-33}
     \mC_{\bar{3}3} = -\frac{\sqrt{3}}{2}\,(\mC_{11}-\mC_{11}')\,, \quad
     \mC_{6\bar{6}} = \sqrt{\frac{3}{8}}\,(\mC_{11}+\mC_{11}')\,,
  \end{equation}
  as well as the remaining octet-octet tensors:
  \begin{equation}
     \mC_{88} = \frac{\mC_{11} - 3\,\mC_{11}'}{2\sqrt{2}}\,, \quad
     \mC_{88}' = \frac{\mC_{11}' - 3\,\mC_{11}}{2\sqrt{2}}\,.
  \end{equation}
  The tensors $\{ \mC_{11},\, \mC_{88} \}$, $\{ \mC_{11}',\, \mC_{88}' \}$ and $\{\mC_{\bar{3}3},\, \mC_{6\bar{6}} \}$
  are mutually orthogonal.
  The color tensors are invariant under both charge-conjugation operations~(\ref{cc-13-24}--\ref{cc-14-23}), i.e., they carry a positive sign, whereas
  either of the two Pauli symmetries~\eqref{pauli-12} or~\eqref{pauli-34} transforms them into each other: $\mC_{11} \leftrightarrow \mC_{11}'$
  and therefore $\mC_{\bar{3}3} \leftrightarrow -\mC_{\bar{3}3}$.

  Next, we write down the Dirac-color tensors. A scalar tetraquark has 256 linearly independent Dirac tensors
  (see Appendix~A in Ref.~\cite{Wallbott:2019dng} for the full basis construction),
  which together with the two independent color structures amounts to 512 tensors in total:
  \begin{equation}\label{0+tensors}
  \begin{split}
     \phi_1 &= \gamma^5_{\alpha\gamma}\,\gamma^5_{\beta\delta}\,\mC_{11}\,, \\
     \phi_2 &= \gamma^5_{\alpha\delta}\,\gamma^5_{\beta\gamma}\,\mC_{11}'\,, \\
     \phi_3 &= \gamma^\mu_{\alpha\gamma}\,\gamma^\mu_{\beta\delta}\,\mC_{11}\,, \\
     \phi_4 &= \gamma^\mu_{\alpha\delta}\,\gamma^\mu_{\beta\gamma}\,\mC_{11}'\,, \\
     \phi_5 &= (\gamma_5 C)_{\alpha\beta}\,(C^T \gamma_5)_{\gamma\delta}\,\mC_{\bar{3}3}\,, \\
     \phi_6 &= (\gamma^\mu C)_{\alpha\beta}\,(C^T \gamma^\mu)_{\gamma\delta}\,\mC_{\bar{3}3}\,, \\
            &\;\; \vdots
  \end{split}
  \end{equation}
  Here the Greek subscripts are Dirac indices only and we suppressed the color indices as well as those for the full Dirac-color amplitudes $\phi_i$;
  they are always understood to be in the standard order $\alpha\beta\gamma\delta$ or $ABCD$.

  The tensors in~\eqref{0+tensors} correspond to the dominant `physical' two-body clusters
  in terms of mesons and diquarks:
  $\phi_1$ and $\phi_2$ describe configurations with two pseudoscalar mesons in (13)(24) and (14)(23),
  $\phi_3$ and $\phi_4$ represent vector-vector meson configurations,
  $\phi_5$ a (12)(34) diquark-antidiquark system with two scalar diquarks ($SS$)
  and $\phi_6$ one with two axialvector diquarks ($AA$).
  They are part of a Fierz-complete subset of 16 Dirac tensors that do not depend on any relative momentum,
  i.e., the `$s$ waves' which do not carry orbital angular momentum.
  In principle one could continue the list until the basis is complete.

  Similarly, an axialvector tetraquark with $J^P=1^+$ has $3\times 256 = 768$ linearly independent Dirac tensors (48 of which are $s$ waves),
  which amounts to 1536 Dirac-color tensors in total~\cite{Wallbott:2019dng}:
  \begin{equation}\label{1+tensors}
  \begin{split}
     \psi_1^\pm &= (\gamma^5_{\alpha\gamma}\,\gamma^\mu_{\beta\delta} \pm \gamma^\mu_{\alpha\gamma}\,\gamma^5_{\beta\delta})\,\mC_{11}\,, \\
     \psi_2^\pm &= (\gamma^5_{\alpha\delta}\,\gamma^\mu_{\beta\gamma} \pm \gamma^\mu_{\alpha\delta}\,\gamma^5_{\beta\gamma})\,\mC_{11}'\,, \\
     \psi_3 &= \varepsilon^{\mu\nu\rho\sigma}\,\hat{P}^\nu\,\gamma^\rho_{\alpha\gamma}\,\gamma^\sigma_{\beta\delta}\,\mC_{11}\,, \\
     \psi_4 &= \varepsilon^{\mu\nu\rho\sigma}\,\hat{P}^\nu\,\gamma^\rho_{\alpha\delta}\,\gamma^\sigma_{\beta\gamma}\,\mC_{11}'\,, \\
     \psi_5 &= (\gamma_5 C)_{\alpha\beta}\,(C^T \gamma^\mu)_{\gamma\delta}\,\mC_{\bar{3}3}\,, \\
     \psi_6 &= (\gamma^\mu C)_{\alpha\beta}\,(C^T \gamma_5)_{\gamma\delta}\,\mC_{\bar{3}3}\,, \\
     \psi_7 &= \varepsilon^{\mu\nu\rho\sigma}\,\hat{P}^\nu\,(\gamma^\rho C)_{\alpha\beta}\,(C^T \gamma^\sigma)_{\gamma\delta}\,\mC_{\bar{3}3}\,, \\
            &\;\; \vdots
  \end{split}
  \end{equation}
  We suppressed again the indices of the $\psi_i$ and color tensors.
  $\psi_1^\pm$, $\psi_2^\pm$ are pseudoscalar-vector and
  $\psi_3$, $\psi_4$ vector-vector meson configurations,
  $\psi_5$, $\psi_6$ describe scalar-axialvector diquark configurations ($SA$)
  and $\psi_7$ is one with two axialvector diquarks ($AA$).

  In Table~\ref{tab_Bose} we collect the combinations of the Dirac-color tensors $\phi_i$ and $\psi_i$ that carry definite Pauli symmetry.
  For example, a combination of type $\Phi_{ss}$ is fully symmetric under both (12) and (34),
  $\Phi_{sa}$ is symmetric under (12) but antisymmetric under (34), etc.
  Table~\ref{tab_CC} lists the transformation properties of the $\phi_i$ and $\psi_i$ under charge conjugation.

  The symmetries of the Dirac-color tensors dictate the symmetries of the remaining flavor tensors.
  As an example, consider the leading Dirac-color-flavor tensors $(\gamma_5 C)\,\mC_{\bar{3}3}\,\mF_0$
  for a scalar diquark and $(\gamma^\mu C)\,\mC_{\bar{3}3}\,\mF_1$ for an axialvector diquark.
  Here, $\mF_0=[ud]$ is the $I=0$ flavor wave function and $\mF_1$ denotes the isospin triplet consisting of $uu$, $\{ud\}$ and $dd$, where
  $[\dots]$ means antisymmetrization and $\{\dots\}$ symmetrization.
  The color tensor $\mC_{\bar{3}3}$ is Pauli-antisymmetric and the Dirac and flavor parts are either both antisymmetric or both symmetric,
  which ensures the antisymmetry of the total wave function. As a consequence, scalar diquarks carry isospin $I=0$ and axialvector diquarks $I=1$.
  One can also construct diquarks with reversed isospin by
  including Pauli-antisymmetric momentum prefactors $p_1^2-p_2^2$ (in analogy to quantum-number exotic mesons), but those
  prefactors would induce a strong suppression of the amplitudes.
  Following this argument, we restrict ourselves to dressing functions
  $f_i(\dots)$ in Eq.~\eqref{general-decomposition2} that are even under all symmetries, so that
  the symmetry properties are carried by the basis tensors $\tau_i$ alone.

  The flavor wave functions depend on the explicit quark content ($qq\overline{qq}$, $cc\overline{qq}$ or $cq\overline{qc}$) and determine the isospin of the state.
  They also transform under the respective symmetries.
   Then, for a system which has Pauli symmetry, the Dirac-color amplitudes in Table~\ref{tab_Bose} must be combined
  with appropriate flavor wave functions $\mF$ to form a totally antisymmetric tensor:
  \begin{equation} \label{DCF-full}  \renewcommand{\arraystretch}{1.4}
     \begin{array}{l}
     \Phi_{ss}\,\mF_{aa}\,, \\
     \Phi_{aa}\,\mF_{ss}\,,
     \end{array}\qquad
     \begin{array}{l}
     \Psi_{ss}\,\mF_{aa}\,, \\
     \Psi_{aa}\,\mF_{ss}\,,
     \end{array}\qquad
     \begin{array}{l}
     \Psi_{sa}\,\mF_{as}\,, \\
     \Psi_{as}\,\mF_{sa}\,.
     \end{array}
  \end{equation}
  If $C$ parity does not apply (such as for $cc\overline{qq}$), these are  the final Dirac-color-flavor wave functions;
  otherwise ($qq\overline{qq}$) one must find appropriate combinations which also have definite $C$ parity.
  If the system has no Pauli symmetry and is only subject to charge-conjugation invariance in one channel
  ($cq\overline{qc}$), one can combine the $\phi_i$ and $\psi_i$ directly with the flavor tensors to obtain total wave functions with definite $C$ parity.

  \begin{table}[t] \renewcommand{\arraystretch}{1.2}
 	\begin{tabular}{c @{\qquad} c @{\qquad} c @{\qquad} c}
 		                      &                       &                     &                           \\[1mm]  \hline \rule{-1.3mm}{0.4cm}
 		$\Phi_{ss}$           & $\Phi_{aa}$           & $\Phi_{sa}$         & $\Phi_{as}$               \\[1mm]  \hline \rule{-1.3mm}{0.4cm}
 		$\phi_1 + \phi_2$       & $\phi_1 - \phi_2$       &                       &                             \\
        $\phi_3 + \phi_4$       & $\phi_3 - \phi_4$       &                       &                             \\
        $\phi_5$                & $\phi_6$                &                       &                             \\[1mm]  \hline \rule{-0.0mm}{0.4cm}

 		$\Psi_{ss}$           & $\Psi_{aa}$           & $\Psi_{sa}$         & $\Psi_{as}$               \\[1mm]  \hline \rule{-1.3mm}{0.4cm}
        $\psi_1^+ + \psi_2^+$   & $\psi_1^+ - \psi_2^+$   & $\psi_1^- - \psi_2^-$ & $\psi_1^- + \psi_2^-$       \\
                                & $\psi_7$                & $\psi_3 - \psi_4$     & $\psi_3 + \psi_4$           \\
                                &                         & $\psi_5$              & $\psi_6$
 	\end{tabular}
 	\caption{Combinations of Dirac-color tensors~(\ref{0+tensors}--\ref{1+tensors}) with definite Pauli symmetry.
 		\label{tab_Bose}}
 \end{table}

  \begin{table}[t] \renewcommand{\arraystretch}{1.2}
 	\begin{tabular}{l @{\;\;} | @{\;\;} r @{\;\;} r @{\;\;} r @{\;\;} r @{\;\;} r @{\;\;} r @{\;\;} r @{\;\;} r @{\;\;} r }
 		            & $\phi_i$ & $\psi_{1,2}^+$  & $\psi_7$  & $\psi_1^-$  & $\psi_2^-$  & $\psi_3$  & $\psi_4$  & $\psi_5$  & $\psi_6$  \\[1mm]  \hline \rule{-1.1mm}{0.4cm}
 		$(13)(24)$  & $\phi_i$ & $-\psi_{1,2}^+$ & $-\psi_7$ & $-\psi_1^-$ & $\psi_2^-$  & $\psi_3$  & $-\psi_4$ & $\psi_6$  & $\psi_5$  \\
 		$(14)(23)$  & $\phi_i$ & $-\psi_{1,2}^+$ & $-\psi_7$ & $\psi_1^-$  & $-\psi_2^-$ & $-\psi_3$ & $\psi_4$  & $-\psi_6$ & $-\psi_5$ \\
    \end{tabular}
 	\caption{Charge-conjugation transformations of the Dirac-color tensors~(\ref{0+tensors}--\ref{1+tensors}) according to the l.h.s. of
             Eqs.~\eqref{cc-13-24} and~\eqref{cc-14-23}.
 		\label{tab_CC}}
 \end{table}

  The resulting Dirac-color-flavor tensors that can be constructed in this way are listed 
  in Table~\ref{tab_0+1+}, together with the isospin and $C$ parities they carry.
  In the following we discuss the construction for different quark content in detail.

  \begin{table*}[t] \renewcommand{\arraystretch}{1.4}
 	\begin{tabular}{l @{\qquad} l @{\qquad} l @{\qquad} l @{\qquad} l}
                                &                      &             &                           &                                \\[1mm]  \hline \rule{-1.3mm}{0.48cm}	
                                & $J^{P(C)}$           & $I$         & Tensors                    & Physical components           \\[1mm]  \hline \rule{-1.3mm}{0.4cm}		
        $nn\overline{nn}$       & $0^{++}$             & $0$         & $\Phi_{ss}\,\mF_{aa}$    & $\{\pi\pi,\, \eta\eta \}$, $\{\rho\rho,\,\omega\omega\}$, $SS$     \\
 		                        &                      & $0,1,2$     & $\Phi_{aa}\,\mF_{ss}$    & $\{\pi\pi, \,\eta\eta,\, \eta\pi \}$, $\{\rho\rho,\,\omega\omega, \, \rho\omega\}$, $AA$   \\
        $cc\overline{cc}$       & $0^{++}$             & $0$         & $\Phi_{aa}\,\mF_{ss}$    & $\eta_c\,\eta_c$, $J/\psi\,J/\psi$, $AA$     \\
        $cc\overline{nn}$       & $0^{+}$              & $1$         & $\Phi_{aa}\,\mF_{ss}$    & $DD$, $D^\ast D^\ast$, $AA$     \\
        $cc\overline{ss}$       & $0^{+}$              & $0$         & $\Phi_{aa}\,\mF_{ss}$    & $D_s D_s$, $D_s^\ast D_s^\ast$, $AA$     \\
        $cn\overline{nc}$       & $0^{++}$             & $0,1$       & $\Phi_1\,\mF_0$, $\Phi_1\,\mF_1$    & $DD$, $\eta_c \,\{\eta,\pi\}$, $D^\ast D^\ast$, $J/\psi\,\{\omega,\rho\}$, $SS$, $AA$     \\
        $cs\overline{sc}$       & $0^{++}$             & $0$         & $\Phi_1\,\mF_0$    & $D_s D_s$, $\eta_c \eta_s$, $D_s^\ast D_s^\ast$, $J/\psi\,\phi$, $SS$, $AA$   \\[1mm]  \hline \rule{-1.3mm}{0.48cm}		
        $nn\overline{nn}$       & $1^{+-}$             & $0$         & $\Psi_{ss}\,\mF_{aa}$    & $\{\pi\rho,\, \eta\omega \}$     \\
 		                        &                      & $0,1,2$     & $\Psi_{aa}\,\mF_{ss}$    & $\{\pi\rho, \,\pi\omega,\, \eta\rho,\,\eta\omega \}$,  $AA$   \\
 		                        &                      & $1$         & $\Psi_-$                  & $\{\pi\rho, \,\pi\omega,\, \eta\rho \}$, $SA$   \\
 		                        & $1^{++}$             & $1$         & $\Psi_+$                 & $\rho\omega$, $SA$   \\
        $cc\overline{cc}$       & $1^{+-}$             & $0$         & $\Psi_{aa}\,\mF_{ss}$    & $J/\psi\,\eta_c$, $AA$     \\
        $cc\overline{nn}$       & $1^{+}$              & $0$         & $\Psi_{as}\,\mF_{sa}$    & $DD^\ast$, $D^\ast D^\ast$, $AS$     \\
                                &                      & $1$         & $\Psi_{aa}\,\mF_{ss}$    & $DD^\ast$, $AA$     \\
        $cc\overline{ss}$       & $1^{+}$              & $0$         & $\Psi_{aa}\,\mF_{ss}$    & $D_s D_s^\ast$, $AA$     \\
        $cn\overline{nc}$       & $1^{++}$             & $0,1$       & $\Psi_1\,\mF_0$, $\Psi_1\,\mF_1$    & $DD^\ast$, $J/\psi \,\{\omega,\rho\}$, $SA$     \\
                                & $1^{+-}$             & $0,1$       & $\Psi_2\,\mF_0$, $\Psi_2\,\mF_1$    & $DD^\ast$, $J/\psi \,\{\eta,\pi\}$, $\eta_c\,\{\omega,\,\rho\}$, $D^\ast D^\ast$, $SA$, $AA$     \\
        $cs\overline{sc}$       & $1^{++}$             & $0$         & $\Psi_1\,\mF_0$    & $D_s D_s^\ast$, $J/\psi\,\phi$, $SA$ \\
                                & $1^{+-}$             & $0$         & $\Psi_2\,\mF_0$    & $D_s D_s^\ast$, $J/\psi\,\eta_s$, $\eta_c\,\phi$, $D_s^\ast D_s^\ast$, $SA$, $AA$ \\
 	\end{tabular}
 	\caption{Dirac-color-flavor wave functions for $J^P=0^+$ (top) and $J^P=1^+$ (bottom) obtained from the tensors in Eqs.~(\ref{0+tensors}--\ref{1+tensors}), together with their isospin and physical interpretation. The explicit flavor wave functions depend on the quark content and are explained in the text.
 		\label{tab_0+1+}}
 \end{table*}

 \newpage

  {\tiny$\blacksquare$} $nn\overline{nn}$: A system made of four light quarks $n \in \{u,d\}$ satisfies all symmetries~(\ref{pauli-12}--\ref{cc-14-23}).
  From $\mathbf{2}\otimes \mathbf{2} \otimes \mathbf{\bar{2}} \otimes \mathbf{\bar{2}}=$ $(\mathbf{1}_a \oplus \mathbf{3}_s) \otimes (\mathbf{1}_a \oplus \mathbf{3}_s)$
  one obtains 16 flavor wave functions with definite Pauli symmetry, which can be arranged into isospin multiplets:
  \begin{equation}
  \begin{split}
     \mF_{aa} \quad &\dots \quad I=0\,, \\
     \mF_{ss} \quad &\dots \quad I = 0, 1, 2\,, \\
     \mF_{as} \quad &\dots \quad I=1\,, \\
     \mF_{sa} \quad &\dots \quad I=1\,.
  \end{split}
  \end{equation}
  For example, $\mF_{aa} = [ud][\bar{u}\bar{d}]$ is the $I=0$ flavor wave function typically associated with the $\sigma$ meson.
  $\mF_{aa}$ and the neutral members of the multiplets for $\mF_{ss}$ have positive $C$ parity under any of the two charge-conjugation operations,
  whereas the remaining ones transform into each other: $\mF_{as} \leftrightarrow \mF_{sa}$ under (13)(24) and $\mF_{as} \leftrightarrow -\mF_{sa}$
  under (14)(23).


  The $C$ parities of the full wave functions $\Phi_{ss}\,\mF_{aa}$ and $\Phi_{aa}\,\mF_{ss}$ from Eq.~\eqref{DCF-full} can be read off from Table~\ref{tab_CC}.
  Since all $\phi_i$ as well as $\mF_{aa}$ and $\mF_{ss}$ have positive $C$ parity,
  the same is true for their combinations and thus they describe $I(0^{++})$ states, where the isospin depends on the flavor wave functions.
  Similarly, $\psi_{1,2}^+$ and $\psi_7$ have negative $C$ parity so that the combinations $\Psi_{ss}\,\mF_{aa}$, $\Psi_{aa}\,\mF_{ss}$ describe $I(1^{+-})$ states.
  The remaining wave functions $\Psi_{sa}\,\mF_{as}$ and $\Psi_{as}\,\mF_{sa}$ carry $I=1$ due to their flavor parts;
  their combinations with definite $C$ parity are given by
  \begin{equation} \renewcommand{\arraystretch}{1.4}
     \Psi_\pm = \left\{ \begin{array}{l} (\psi_1^--\psi_2^-)\,F_{as} \mp (\psi_1^-+\psi_2^-)\,F_{sa}\,, \\
                                       (\psi_3-\psi_4)\,F_{as} \pm (\psi_3+\psi_4)\,F_{sa}\,, \\
                                       \psi_5\,F_{as} \pm \psi_6\,F_{sa}\,,
               \end{array}\right.
  \end{equation}
  where $\Psi_\pm$ corresponds to $I(J^{PC})=1(1^{+\pm})$.

  The final Dirac-color-flavor wave functions satisfy all four constraints in Eqs.~(\ref{pauli-12}--\ref{cc-14-23}).
  Each has an interpretation in terms of a meson-meson or \mbox{$dq$-$\conjg{dq}$} configuration. The corresponding physical
  two-body systems are listed in the right column of Table~\ref{tab_0+1+}.


  \smallskip

  {\tiny$\blacksquare$} $cc\overline{cc}$: For an all-charm state there is only one flavor wave function $\mF_{ss} = cc\overline{cc}$
  which is symmetric under any quark exchange, invariant under both charge-conjugation operations and carries $I=0$.
  As a consequence, only the combinations $\Phi_{aa}\,\mF_{ss}$ and $\Psi_{aa}\,\mF_{ss}$ in Eq.~\eqref{DCF-full} survive
  and produce states with quantum numbers $0(0^{++})$ and $0(1^{+-})$, respectively.
  Note that due to the symmetry of $\mF_{ss}$
  there is no overlap with the `$\sigma-$like' components $\Phi_{ss}$ in the $nn\overline{nn}$ system.

  \smallskip

  {\tiny$\blacksquare$} $cc\overline{nn}$: For an open-charm system, the light quarks can be arranged into
  an isosinglet $[\bar{u}\bar{d}]$ and an isotriplet $\bar{u}\bar{u}$, $\{\bar{u}\bar{d}\}$, $\bar{d}\bar{d}$.
  Thus there are four flavor wave functions:
  \begin{equation}
  \begin{split}
     \mF_{sa} \quad &\dots \quad I=0\,, \\
     \mF_{ss} \quad &\dots \quad I=1\,.
  \end{split}
  \end{equation}
  In this case charge conjugation symmetry does not apply, so that the final wave functions are
  $\Phi_{aa}\,\mF_{ss}$, $\Psi_{aa}\,\mF_{ss}$, $\Psi_{as}\,\mF_{sa}$ with quantum numbers $1(0^+)$, $1(1^+)$ and $0(1^+)$, respectively.
  Note that this does not produce a $0(0^+)$ state since there is no entry $\Phi_{as}$ in Table~\ref{tab_Bose}.

  \smallskip

  {\tiny$\blacksquare$} $cc\overline{ss}$: Here the only flavor wave function is $\mF_{ss}$, however with $I=0$,
  so that the combinations $\Phi_{aa}\,\mF_{ss}$ and $\Psi_{aa}\,\mF_{ss}$ produce the quantum numbers $0(0^+)$ and $0(1^+)$.
  Compared to the previous case, going from light to strange quarks for the same quantum numbers therefore leads to quite different wave-function components.

  \smallskip

  {\tiny$\blacksquare$} $cn\overline{nc}$: In the hidden-charm system Bose symmetry does not apply.
  Instead, the flavor wave functions
  \begin{equation}\label{hidden-charm-flavor}
  \begin{split}
     \mF_0 &= c (u\bar{u}+d\bar{d}) \bar{c} \quad\quad\quad \;\,  \dots \quad I=0\,, \\
     \mF_1 &= \left\{\begin{array}{c} cu\bar{d}\bar{c} \\ c (u\bar{u}-d\bar{d}) \bar{c} \\ c d\bar{u}\bar{c} \end{array}  \right\} \quad  \dots \quad I=1
  \end{split}
  \end{equation}
  have positive $C$ parity in (14)(23).
  The $C$ parities of the full wave functions $\{\phi_i, \psi_i\} \times \{\mF_0, \mF_1\}$ can then be read off
  from the bottom line of Table~\ref{tab_CC}: The Dirac-color amplitudes
  \begin{equation}
  \begin{split}
     \Phi_1 &= \{ \phi_1, \, \phi_2, \, \phi_3, \, \phi_4, \, \phi_5,\, \phi_6 \}\,, \\
     \Psi_1 &= \{ \psi_1^-,\, \psi_4, \, \psi_5-\psi_6 \} \,, \\
     \Psi_2 &= \{ \psi_1^+, \, \psi_2^+ \pm \psi_2^-, \, \psi_3, \, \psi_5+\psi_6, \, \psi_7 \}
  \end{split}
  \end{equation}
  correspond to the quantum numbers $I(0^{++})$, $I(1^{++})$ and $I(1^{+-})$, respectively,
  where the isospin $I=0,1$ depends on the flavor wave function~\eqref{hidden-charm-flavor}.

  \smallskip

  {\tiny$\blacksquare$} $cs\overline{sc}$: Here the situation is analogous except there is only one
  flavor wave function $\mF_0 = cs\bar{s}\bar{c}$ with $I=0$.

  \smallskip

  Up to this point we have made two approximations in the construction of the Bethe-Salpeter amplitude.
  Instead of the complete tensor basis, we have restricted ourselves to the dominant tensors in Eqs.~(\ref{0+tensors}--\ref{1+tensors}),
  and we only considered dressing functions $f_i$ that are even under the relevant symmetry operations.
  Those dressing functions, however, still depend on ten independent variables: the total momentum $P^2 = (p_1 + p_2 + p_3 + p_4)^2 = -M^2$,
  where $M$ is the mass of the tetraquark, plus nine variables that are dynamical.
  This is what makes a full numerical solution practically impossible.

  In the final step we therefore assume that the momentum dependence of the $f_i$ is mainly carried by the symmetric variable
  \begin{equation}
     \mS_0 = \frac{1}{4}\left(p_1^2 + p_2^2 + p_3^2 + p_4^2 + \frac{M^2}{4}\right),
  \end{equation}
  and that the remaining angular dependencies can be well described by two-body intermediate poles
  corresponding to the physical components listed in Table~\ref{tab_0+1+}. Such a behavior has been found
  directly in the dynamical solution of the four-body equation for light scalar tetraquarks~\cite{Eichmann:2015cra}
  and it is also the approximation employed in Ref.~\cite{Wallbott:2019dng}.
  We therefore assume
  \begin{equation}
     f_i(\dots) \approx f_i(\mS_0)\,\mP(m_1, m_2)_{ab,cd}\,,
  \end{equation}
  where the pole factors depend on the respective tensor component:
  \begin{equation*}
     \mP(m_1,m_2)_{ab,cd} = \frac{1}{(p_a+p_b)^2 + m_1^2}\,\frac{1}{(p_c+p_d)^2 + m_2^2}\,.
  \end{equation*}
  For example, for a $cn\overline{nc}$ system with $I=0$, the pole factors corresponding to the $\phi_i$ in Eq.~\eqref{0+tensors} are
  \begin{equation}
  \begin{split}
     &\mP(m_D,m_D)_{13,24} \,, \\
     &\mP(m_{\eta_c},m_\eta)_{14,23} \,, \\
     &\mP(m_{D^\ast},m_{D^\ast})_{13,24} \,, \\
     &\mP(m_{J/\psi},m_{\omega})_{14,23} \,, \\
     &\mP(m_S,m_S)_{12,34} \,, \\
     &\mP(m_A,m_A)_{12,34} \,,
  \end{split}
  \end{equation}
  where we compute all meson and diquark masses from their two-body Bethe-Salpeter equations.
  Note that these pole factors are invariant under the symmetry operations~(\ref{pauli-12}--\ref{cc-14-23})
  that are relevant for the particular system (in the $cn\overline{nc}$ example
  only charge conjugation in (14)(23) applies), so we can also attach them directly to the tensors~(\ref{0+tensors}--\ref{1+tensors}).
  As a result, only the dependence on the symmetric variable $\mS_0$ remains dynamical.

  The $f_i(\mS_0)$ are therefore the amplitudes
  that are determined in the solution of the four-body equation shown in Fig.~\ref{fig_q4_bse}, together with the tetraquark mass $M$.
  By switching off individual tensor components we can determine their strengths and isolate the dominant contributions.
  This completes our construction of the tetraquark Bethe-Salpeter amplitude.

%
%
%

\end{appendix}

\end{document}